\documentclass[usenatbib, usedcolumn]{mn2e}
\usepackage{times}
\usepackage{multirow}
\usepackage{amsmath}
\usepackage{graphicx}
\usepackage[section,below]{placeins}   % gives a \FloatBarrier to force all floats to appear

\title{Clocks around Sgr A*}

\author[Ang\'elil \& Saha]
  {Raymond Ang\'elil$^1$\thanks{rangelil@physik.uzh.ch}, Prasenjit Saha$^2$\\
  $^1$Institut f\"ur Rechnergest\"utztewissenschaften, Universit\"at Z\"urich,
Winterthurerstrasse 190, 8057 Z\"urich, Switzerland \\
  $^2$Physik-Institut, Universit\"at Z\"urich,
Winterthurerstrasse 190, 8057 Z\"urich, Switzerland }

\date{\today}

\begin{document}

\maketitle

\begin{abstract}
The S~stars near the Galactic centre and any pulsars that may be on
similar orbits, can be modelled in a unified way as clocks orbiting a
black hole, and hence are potential probes of relativistic effects,
including black hole spin.  The high eccentricities of many S~stars
mean that relativistic effects peak strongly around pericentre; for
example, orbit precession is not a smooth effect but almost a kick at
pericentre.  We argue that concentration around pericentre will be an
advantage when analysing redshift or pulse-arrival data to measure
relativistic effects, because cumulative precession will be drowned
out by Newtonian perturbations from other mass in the Galactic-centre
region.  Wavelet decomposition may be a way to disentangle
relativistic effects from Newton perturbations.  Assuming a plausible
model for Newtonian perturbations on S2, relativity appears to be
strongest in a two-year interval around pericentre, in wavelet modes
of timescale $\approx 6$~months.
\end{abstract}

\section{Clocks as probes of gravity}

An orbiting clock as a probe of general relativity is familiar from
binary pulsars \citep{1994RvMP...66..711T,Kramer} and from global
navigation satellites \citep{2003LRR.....6....1A}.  In the coming
years, a new class of objects may join these.

The milliparsec region of the Galactic centre is home to a compact
mass of $\sim 4\cdot 10^6M_{\odot}$ at Sgr~A*, presumably a black
hole.  This is known from a population of stars which orbit it at
speeds up to a few percent of light, as shown by astrometric and
spectroscopic observations \citep{originalS2, gc_distance, earliest,
  spec1, keck, gillessenMonitors, ghez, meyer}. Dozens of these
`S'-stars have been observed, and it is expected that many others with
orbits even closer to Sgr~A* await discovery by the next generation of
telescopes such as the European Extremely Large Telescope (E-ELT).  An
even more exciting prospect is the possibility of pulsars near the
black hole.  A pulsar has recently been discovered in the region
\citep{2013ApJ...775L..34R}, and population models argue that there
should be a few pulsars with periods $<1\,\rm yr$ and observable with
the Square Kilometre Array \citep{pulsarsInGC, evenMoarPulsarsInGC,
  morePulsarsInGC, pulsarSearch}.
Since the gravitational radius of the black hole is
\begin{equation} \label{eq:gravradius}
   \frac{GM}{c^2} \simeq 20\;\hbox{light-sec,}
\end{equation}
and recalling that a parsec is $1.0\times10^8$ light-seconds, we see
that the S~stars are at $r\sim10^4$ in relativistic units.  This make
their orbits the most relativistic of all known ballistic orbits, more
than any known binary system, and far more than Mercury or artificial
satellites \citep[see, for example, Figure~2 in][]{paper2} and
provides incentive to search for relativistic effects. 

Progress has been made to directly observe the event horizon silhouette\citep{2010evn..confE..53D}, and these two kinds of observations could potentially complement each other\citep{2014ApJ...784....7B}.

\begin{figure}
\includegraphics[scale = 0.45]{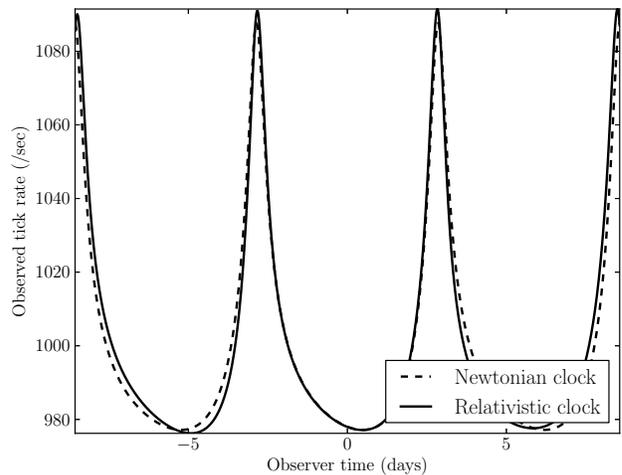}
\caption{Illustration of the basic scenario: an observer watches a
clock in a nearly Keplerian orbit around black hole, but relativity
changes both the clock's orbit, and the paths that the signals take to
reach the observer.  The orbits are initialised at zero proper time
and then integrated forward and back, but observer time lags by about
a day, because of the placement of the observer.  Note how the tick
rates are very similar until pericentre passage, but then the
relativistic orbit appears to get ahead a little bit --- that is
pericentre precession.  Spin effects are also included in the
calculation, but too small to see at the resolution of this
figure. The eccentricity of this orbit is $e=0.6$, the semi-major axis is $a=0.06\,$mpc, and the inclination $I=45^\circ$. For this and all other numerical results in this paper, we set the black hole mass to $20$~light seconds, cf. equation \eqref{eq:gravradius} .}\label{fig:funclox} \end{figure}

An S~star, or a pulsar on a similar orbit, can be considered as a
clock in orbit around a black hole.  The clock moves on a time-like
geodesic and ticks at equal intervals of its proper time
$\tau_c=n\nu_c$.  With each tick, the clock sends out photons in all
directions on null geodesics.  Some of these photons reach an
observer, who records their arrival times as $t_a(n)$.  The observer
can also choose to calculate the frequency by taking the derivative of
the arrival times with respect to the proper time of emission:
\begin{equation}\label{crux}
  \nu_a = \frac1{t_a(n+1)-t_a(n)} =
  \nu_c \left(\frac{dt_a}{d\tau_e}\right)^{-1}.
\end{equation}
Figure~\ref{fig:funclox} shows an example of what might be measured.
For pulsars, the ticks are simply the pulses.  For an S~star, there
are no such discrete ticks, but the clock model still applies, because
\begin{equation}\label{crux2}
   c \, \ln (\nu_c/\nu_a)
\end{equation}
has the interpretation of redshift as usually measured from
spectroscopy\footnote{Redshifts are conveniently stated in km/s, but
in relativity no longer correspond to radial velocities.}.  It is not
essential for the observer to know the intrinsic frequency in advance,
since $\nu_c$ just introduces an additive constant into equation \eqref{crux2}.
The important thing is to be able to calculate $t_a(n)$, for which one
has to to compute time-like geodesics (orbits) and null geodesics
(light paths), and solve the boundary-value problem for null geodesics
from clock to observer.  A moving observer can also be allowed for, if
desired.

Many different relativistic effects are, in principle, measurable from a clock orbiting a black hole.  First, the spacetime around the
black hole dilates the clock time.  Then, every term in the metric
affects both the orbit of the clock and the photons from the clock,
and imprints itself on the observables in its own distinctive way.
The best-known examples are pericentre precession and the Shapiro time
delay; the former concerns orbits while the latter influences light
paths.  Another difference is that precession is cumulative over many
orbits, whereas the Shapiro delay is transient and does not get larger
as one observes more orbits.  In the solar system and for binary
pulsars, both cumulative and transient effects are measurable.  The
circumstances of the Galactic-centre region, however, strongly favour
the transients over the cumulatives for the following reasons:
\begin{enumerate}
\item The orbital periods are long. Cumulative build-up needs multiple
  orbits which takes decades.
\item Orbits of S~stars tend to be highly eccentric, $e=0.9$ being
  typical.  Relativistic effects increase more steeply with small
  radius and high velocity than classical effects, and hence relativity is strongest
  around pericentre passage.
\item The extended stellar system will contribute significant noise,
  hampering in particular searches which rely on build-up over long
  time scales. Whereas it may be possible to disentangle transient
  effects from noise over short time scales because the time
  dependence of the former is well understood.
\end{enumerate}

With a full 4-dimensional relativistic treatment, this paper performs
numerical experiments - computing arrival times $t_a$ - to gain
insight into transient relativistic behaviour on S~Star
redshifts. Section \ref{sec:context} discusses the more familiar tests
of the Kerr metric and discusses a few examples of transient effects,
section \ref{sec:scalings} formalises our redshift-calculating method,
and discusses how the different effects scale with orbital
period. Section \ref{sec:results} calculates these effects for mock
S~Star orbits. Finally, in section \ref{sec:wavelets} we propose a
novel strategy based on wavelet decomposition which may help separate
relativistic behaviour from Newtonian noise.

%\FloatBarrier

\section{Familiar effects from Kerr}\label{sec:context}

There are multiple relativistic effects which grow over many orbits. This has
been essential to observing them in artificial satellites, planets or
pulsars. Here we list some of the well-known ones.
\subsection{Cumulative}
\begin{enumerate}
\item The expected relativistic orbital precession has been discussed
extensively in the context of S~stars \citep[e.g.,][]{rubilar,
  Merritt, will2, new} and pulsars \citep[e.g.,][]{wex}.  Relativity
gives several contributions to the precession. The strongest cumulative relativistic effect comes from the first Schwarzschild contribution, resulting in a perihelion shift 
\begin{equation}\label{mercury}
\Delta \omega = \frac{6\pi}{a(1-e^2)}
\end{equation}
per orbit\footnote{In this paper, all lengths are measured in units
  of the gravitational radius $GM/c^2$.}. 
  \item There is another contribution to the precession if the black hole has
internal angular momentum. This is characterised by a spin parameter; or angular
momentum per unit mass $s$. Bodies near the black hole experience
frame dragging in the spin direction. The precessional effect due to this is 
\begin{equation}\label{fd_precession}
\Delta \phi = -8\pi s \, [a(1-e^2)]^{-3/2}
\end{equation}
per orbit.  The phenomenon of frame dragging has been first observed
only in recent years, by using laser ranging to accurately determine
the orbit of the Lageos satellites and reveal the relativistic effect
of the Earth's spin \citep{lageos,lageos_problems}.  The recently
launched LARES satellite aims to measure the effect to an accuracy of
1\%\citep{lares}.   
\item Two further effects act on the spin of the star or
pulsar.  One is geodetic precession, wherein a vector attached to an
orbiting body moves by (for circular orbits)
\begin{equation}
\Delta \phi = \frac{3\pi}{a}
\end{equation}
per orbit \citep{fliessbach}.  Gravity Probe B has measured this
effect in Earth's gravitational field \citep{gravprobeB_geodetic}.
The parallel transport of a vector along a geodesic is also influenced
by frame-dragging. This is called the Lens-Thirring effect, and was
also detected by Gravity Probe B \citep{gravprobeB_geodetic}. It is
possible that the spin axis of a close-in pulsar be parallel
transported enough to change the pulse profile. Pulse-profile changes
from geodetic precession have been observed in Binary Pulsar systems
\citep{geodetic_precession1,geodetic_precession2,geodetic_precession3,geodetic_precession4},
and could be observed in galactic centre pulsars.

Orbital decay due to
gravitational radiation is another well-known effect, but the time
scales are too slow to be interesting for
Galactic-centre stars.
\end{enumerate}

\begin{figure*}\includegraphics[scale=0.6]{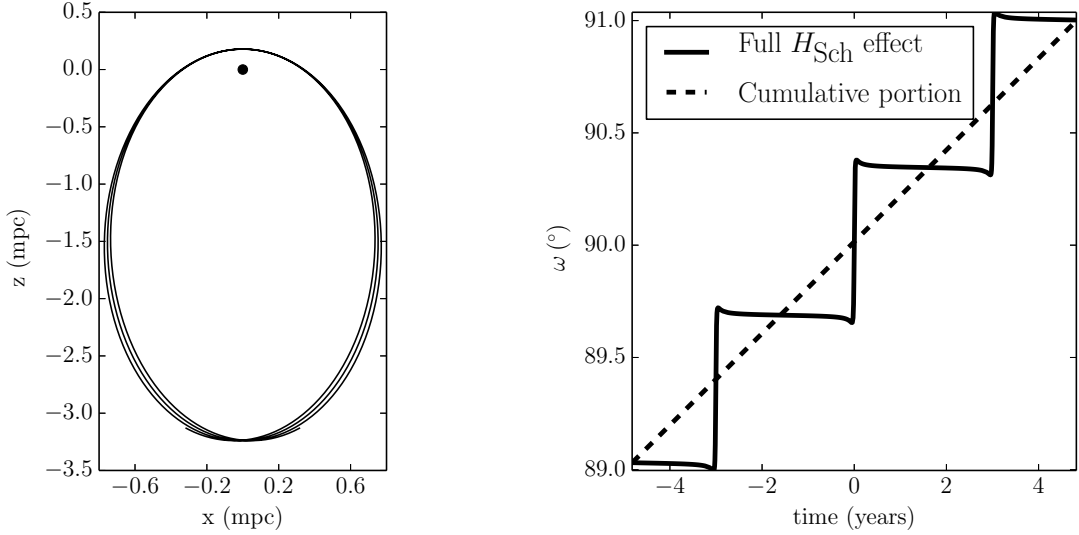}
\caption{Illustration of Schwarzschild precession. The orbit is like a
  1/3-size version of the star S2 \citep{keck}, with semi-major axis
  $a =0.041''$ and eccentricity $e=0.88$, but viewed face-on. We have
  set the distance to the galactic centre at $8.31\,$kpc. Along the
  trajectory we may associate a value for the argument of pericentre
  $\omega$ to the value it would take were the phase space position a
  solution to Kepler's equations. The right panel shows the
  instantaneous argument of pericentre against coordinate time for a
  distant observer. We see here that, for high eccentricities,
  precession is concentrated so strongly around pericentre that it
  looks nearly discrete.  The dashed curve is the well-known
 formula (\ref{mercury}) for the cumulative
  precession.}
\label{fig:orbit_H4}
\end{figure*}

\begin{figure*} \includegraphics[scale=0.6]{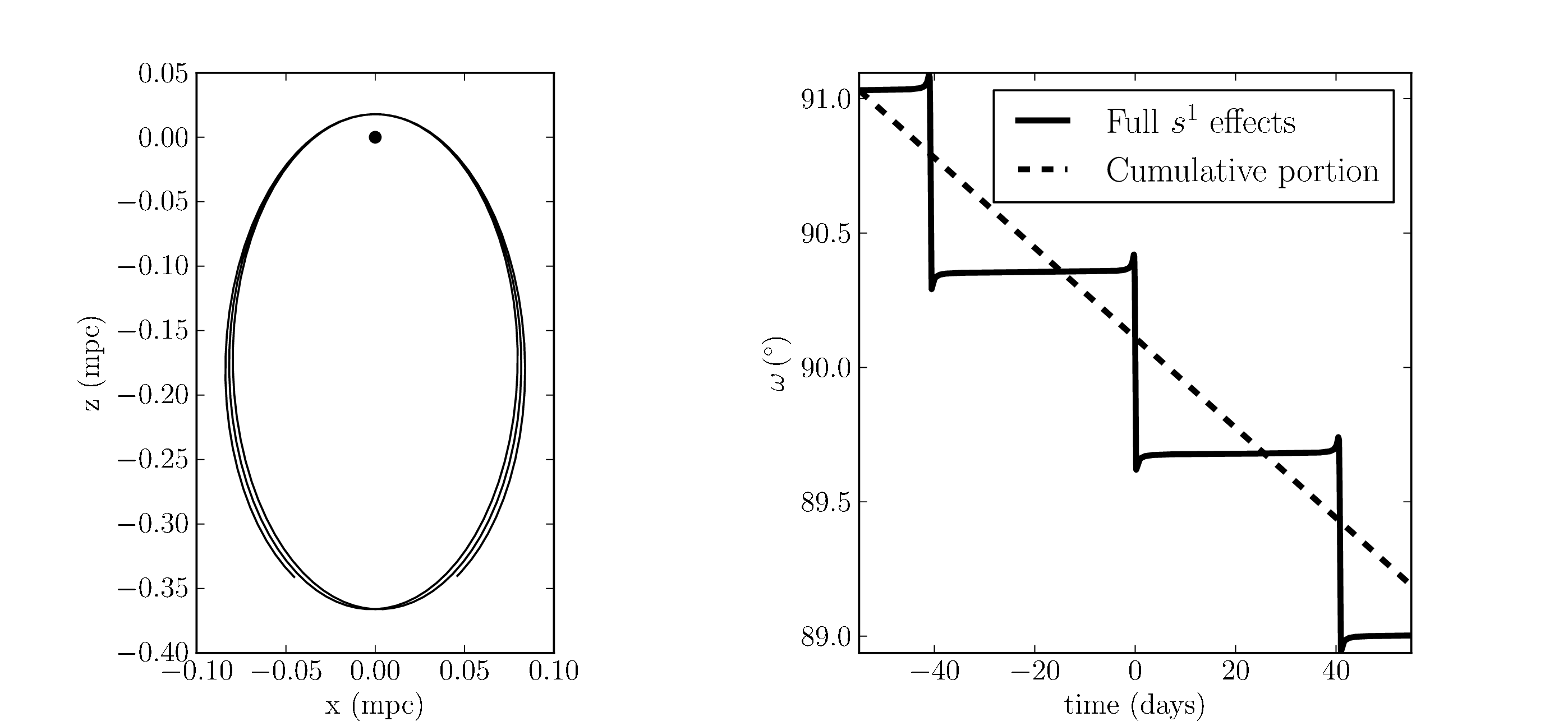}
\caption{Illustration of frame-dragging precession.  Since the effect
is higher order than Schwarzschild precession, a smaller orbit is used
than in Figure~\ref{fig:orbit_H4} in order to make the precession
visible: the orbit is a 1/30-sized version of the star S2 (here with $e=0.88$ and $a=0.0041''$), yet viewed
face-on. The spin is maximal and perpendicular to the orbit; were the
spin direction not perpendicular to the orbital plane, the orbital
plane would also precess about the spin axis. Frame-dragging is even
more strongly concentrated around pericentre than Schwarzschild
precession. The dashed curve is the formula
(\ref{fd_precession}).}  \label{fig:orbit_H5}
\end{figure*}

\subsection{Transients}
Unlike the orbits of satellites, planets or pulsars, in the Galactic
centre, orbital periods are much longer, so accumulating relativistic
signals over many orbits is difficult even though the fields are far
stronger.  Perhaps an even more serious problem is the Newtonian
perturbations due to gas and other stars in the Galactic-centre
region.  So it is interesting to think about transient effects which
occur over a single orbit.  These may be measurable over a short time,
and moreover a predictable time dependence could enable extracting the
signal from the Newtonian background.  In fact, there is a plethora of
such effects, a few of which we describe here.

\begin{enumerate}

\item The strongest relativistic effect is gravitational time
  dilation, one of the basic consequences of the Equivalence
  Principle.  Time is dilated by a factor
  \begin{equation}
  g_{tt}^{-1/2} = 1-\frac{2}{r}
  \end{equation}
  with no effect at this order on the orbit or the light paths.  For
  a highly eccentric orbit, clearly there will be a peak at pericentre.
  GNSS satellites are sensitive to this shift.  For navigation
  demands, it is enough for GNSS satellites to routinely step the
  on-board clock time back, correcting for this effect.  Gravitational
  time dilation has not yet been measured in the galactic centre, but
  is expected to be possible in the near future \citep{zucker}. If
  observed, gravitational time dilation would provide a new test of
  the Einstein Equivalence Principle \citep{paper3}.

\item Lensing effects of gravity on photons travelling to us are
  naturally also transient phenomena.  Astrometric shifts due to
  gravitational lensing have been discussed in the Galactic-centre
  context \citep{bozza}, as have time delays due to a curved
  space-time \citep{paper1}, although none have yet been detected.
  With an impact parameter $b$, the deflection angle of a null ray is
  \begin{equation}
  \Delta \phi = \frac{4}{b}.
  \end{equation}
  This is the leading-order Schwarzschild contribution. The $\sim
  b^{-2}$ effect is also relevant, and enters at the same order as
  the frame-dragging lensing contribution, which we discuss later. The
  extra delay induced in the arrival time of a packet of light
  compared to had it travelled in a straight line is the
  Shapiro delay \citep{shapiro1}, and has been well-tested with the
  \textit{Mariner 9} and \textit{Viking} spacecraft in the solar
  system\citep{shapiro2, Viking1, Viking2} and in binary pulsar
  systems\citep{pulsy-wulsy, 2010Natur.467.1081D}.

\item Underlying every type of orbit precession, there is a fleeting
  contribution which occurs around pericentre, the memory of which is
  not retained by the orbit's shape afterwards.
  Figure~\ref{fig:orbit_H4}, which we shall return to later, shows our first example: the precession
  of the instantaneous pericentre of a highly eccentric orbit.  Far
  from being the smooth effect suggested by equation \eqref{mercury},
  it consists almost of discrete kicks. In the derivation of
  (\ref{mercury}), an oscillatory term crops up beside this
  term. Because this term imparts a momentary perturbation which
  time-averages to zero, it is dropped in textbook
  derivations.\citep{torah, necronomicon, BookOfMormon,BhagavadGita}.
  Analogously, Figure \ref{fig:orbit_H5} shows pericentre precession
  due to frame dragging by the black hole spin, of which
  \eqref{fd_precession} is the average.  These two examples are
  artificial and do not themselves correspond to observable
  quantities, for two reasons.  First, for Figure~\ref{fig:orbit_H5},
  we have dropped lower-order contributions from space curvature so as
  to isolate frame dragging.  Second, the instantaneous pericentre of
  an orbit is defined as the pericentre of a Keplerian orbit with the
  same instantaneous position and momentum \citep[the osculating
    elements, see e.g.][]{1999ssd..book.....M}.  In relativity, the
  instantaneous pericentre therefore becomes gauge-dependent and hence
  is not an observable quantity \citep[cf.][]{preto}.  Nonetheless
  Figures~\ref{fig:orbit_H4} and \ref{fig:orbit_H5} do suggest that
  time-resolved observations could detect relativistic effects over a
  single orbit, especially around pericentre, where relativity is
  strongest and Newtonian perturbations are likely to be at their
  weakest.

\end{enumerate}

In Section~\ref{sec:results} below, we show how these and several
other effects can be readily calculated numerically using a Hamiltonian
formalism, and show various illustrative examples.

\FloatBarrier

\section{Time delays and redshifts in Kerr}\label{sec:results}

\subsection{The Hamiltonian}\label{sec:scalings}

The Hamilton equations for
\begin{equation} \label{superhamiltonian}
   H = {\textstyle\frac12} \, g^{\mu\nu}\, p_\mu \, p_\nu \,,
\end{equation}
are simply the geodesic equations, with the affine parameter
taking on the role of the independent variable.  Since $H$ does not
depend explicitly on the affine parameter, $H$ is constant along a
geodesic.  Proper time is $\sqrt{|H|}$ times the affine parameter,
except for the case of $H=0$, corresponding to null geodesics. 

\def\disp{\displaystyle \vrule width 0pt height 24pt depth 20pt}
\def\desc#1#2{\vbox{\hsize=0.25\hsize
              \parindent=0pt \leftskip=0pt plus1fil \rightskip=\leftskip
              \centerline{\phantom{\bigg|}#1} \medskip
              \centerline{$\Delta t \sim P^{\,#2}$}}}

\begin{table*}
\caption{Hamiltonian terms for orbits and light paths in a Kerr
  spacetime.  The full Hamiltonian \eqref{superhamiltonian} is the sum
  of all the terms in the left column, plus higher-order terms that we
  have not considered.  The middle and right terms group the terms by
  physical effect and scaling of the time delay $\Delta t$ with period
  $P$, as explained in Section~\ref{sec:scalings} Note that we are
  using geometrised units $GM=c=1$ here.  To put $\Delta t$ and $P$ in
  time units, simply multiply by a power of $GM/c^3$ so as to get the
  dimensions right.  \label{tab:Terms}}
\begin{tabular}{|l l||c|c|} \hline 
 & & Orbits & Light paths  \\
 \hline 
&$\disp -\frac{p_t^2}{2}$  & static &\multirow{2}[4]{*}{\desc{R\o mer}{2/3}} \\
\cline{3-1}
& +$\disp\frac{\bf{p}^2}{2}$ & \multirow{2}[6]{*}{\desc{Kepler}{2/3}}&   \\
\cline{4-1}
& $\disp-\frac{p_t^2}{r}$ & & \multirow{2}[4]{*}{\desc{Shapiro}{0}} \\
\cline{3-1}
& $\disp-\frac{(\bf{x}\cdot \bf{p})^2}{r^3}$ &\multirow{2}[6]{*}{\desc{Schwarzschild}{0}} &  \\
\cline{4-1}
&$\disp-\frac{2p_t^2}{r^2}$ & & \multirow{3}[10]{*}{} \\
\cline{3-1}
&$\disp-\frac{2p_t \bf{p}\cdot\left(\bf{s}\times{x} \right)}{r^3}$ & \desc{frame-dragging}{-1/3} & \desc{frame-dragging, spin-squared, Shapiro}{-2/3}\\
\cline{3-1}
&$\disp +\frac{s_\perp^2}{2r^4}\left(\bf{x}\cdot\bf{p} \right)^2 -\frac{1}{2r^4}\frac{\left(\bf{p}\cdot \bf{s}\times \bf{x} \right)^2}{s_\perp^2}$ &\multirow{2}[8]{*} &  \\

%\cline{3-1}
&$\disp-\frac{1}{2r^4}\frac{1-s_\perp^2}{1-s_\parallel^2}\left(\left(\bf{p}\cdot \bf{s}\right)r - \frac{\left(\bf{x}\cdot \bf{s}\right) \left(\bf{x}\cdot\bf{p}\right)}{r} \right)^2$ &\multirow{2}[8]{*}{\desc{Spin (even), Schwarzschild}{-2/3}} &  \\

\cline{4-1}
&$\disp +\frac{p_t^2}{r^3}s_\perp^2 -\frac{4p_t^2}{r^3}$ & & not included \\
\hline
\noalign{\medskip}
\end{tabular}
\end{table*}

In our case, $g^{\mu\nu}$ are the contravariant components of the Kerr metric. The Kerr metric is a vacuum solution to the Einstein Field Equations. This is the appropriate metric to use if we are interested in solving the forward problem for relativistic effects. This will allow us to investigate examples of transient relativistic effects in isolation. In section \ref{sec:wavelets} we treat the more realistic case; we relax the vacuum assumption, and add other S~stars as Newtonian perturbers to the system, and see whether we can uncover transient relativistic effects when the Newtonian noise is significantly large. 

Assuming the orbits and light paths do not go close to the horizon, we
can expand the Hamiltonian in powers of $1/r$. The result is available
in \cite{paper1}.  However because it is convenient to be able to set
the black hole spin direction without having to rotate the observer
and the orbit, we use a slightly different form. The Kerr geometry in
Boyer-Lindquist coordinates necessarily aligns the axis of symmetry of
the coordinate system with the axis of symmetry of the space-time
geometry itself, and it is therefore not possible to disentangle the
preferred direction of the coordinates with that of the spin in these
coordinates. This means we need to first transform to pseudo-Cartesian
coordinates. Table~\ref{tab:Terms} contains the form of the
Hamiltonian which we use here, in pseudo-Cartesian coordinates, and
with the spin promoted to a 3-vector $\mathbf{s} = \left(s_x, s_y,
s_z\right)$.  \citep[An equivalent table is given in][but using
  Boyer-Lindquist variables.]{earth_clox} For convenience, we use the
short-form
\begin{equation}
s_\perp \equiv \frac{\bf{s} \times \bf{x}}{r} \qquad
\textrm{and} \qquad s_\parallel \equiv \frac{\bf{s}\cdot \bf{p}}{r}. 
\end{equation}

Presenting the Hamiltonian in table form allows us to group the terms according
to physical effects on orbits or light paths\footnote{While both orbits and light paths are geodesic in the same metric, the orders at which various terms affect the dynamics differ, due to the different behaviour of their momentum.}. The Kepler/R\o mer terms are classical.  The leading relativistic effect is time dilation, but
as it is not associated with geodesics as such, it does not appear in
the table.  Relativistic terms not depending on the spin parameter $s$
are labelled `Schwarzschild'. Then there are various terms depending
on spin. Of these, the term odd in $s$ gives frame dragging.

We are now ready to use the Hamilton equations corresponding to the Hamiltonian in table~\ref{tab:Terms} to explore the dynamics, and the consequences of the many terms in Table~\ref{tab:Terms}. While \cite{paper2} solves the inverse problem for relativity on S~stars, here we attempt to give a more qualitative picture of exactly how relativity perturbs the orbit and redshifts/arrival-times, in particular for transient effects.

\subsection{Numerical Experiments with S~Stars and Pulsars}

We have already referred to Figure~\ref{fig:funclox}, in passing in
the Introduction.  That figure compares the observable pulse rate from
two cases: (i)~a clock follows a relativistic orbit and the ticks are
conveyed to the observer along null geodesics, and (ii)~the classical
case including Kepler and R\o mer effects, and time dilation.  The
relativistic case includes all terms in Table~\ref{tab:Terms}, other
than the two highest-order ``not included'' terms in the light path.
The orbits are initialised at apocentre with $e=0.6$ and inclination
$I=45^\circ$ with respect to the line of sight, and are integrated forward
and back.  The rest-frame tick rate of the clock is 1000\thinspace Hz
and its orbital period is a week, while the assumed gravitational
radius of the black hole is $GM/c^2=20\,\hbox{light-sec}$
(corresponding to Sgr~A*) --- these choices are only for the sake of
putting axes on the figure and have no physical significance. 

From Figure~\ref{fig:funclox} we can infer that relativity makes the
pericentre precess, but to see more detail we need to extract the
difference between the relativistic and non-relativistic cases.  It is
especially interesting to see what different groups of terms from
Table~\ref{tab:Terms} do to the time-delay and redshift curves.  To
label different cases, let us introduce some shorthand, as follows.
\begin{enumerate}
\item $H_{\rm sch}$ means that Schwarzschild terms but not spin terms
have been included in the orbits, while no relativistic terms have
been included for the light paths. These are the forth and fifth rows in Table~\ref{tab:Terms}.  $H^{\rm sch}$ means that
Shapiro terms have been included for the light paths, while the
orbits are classical. These are the third and fourth rows in Table~\ref{tab:Terms}. 
\item $H_{s}$ ($H^{s}$) means that only the spin $p_t \left(\mathbf{x} \times \mathbf{p}\right)/{r^3}$ term has been
added to the classical terms, and only for orbits (light paths). This term is found in the sixth row in Table~\ref{tab:Terms}.
\item Similarly, $H_{s^2}$ ($H^{s^2}$) means classical plus spin-squared
terms in the orbits (light paths). These terms are the remaining rows of Table~\ref{tab:Terms}.
\end{enumerate}

\begin{figure*}
\includegraphics[scale = 0.5]{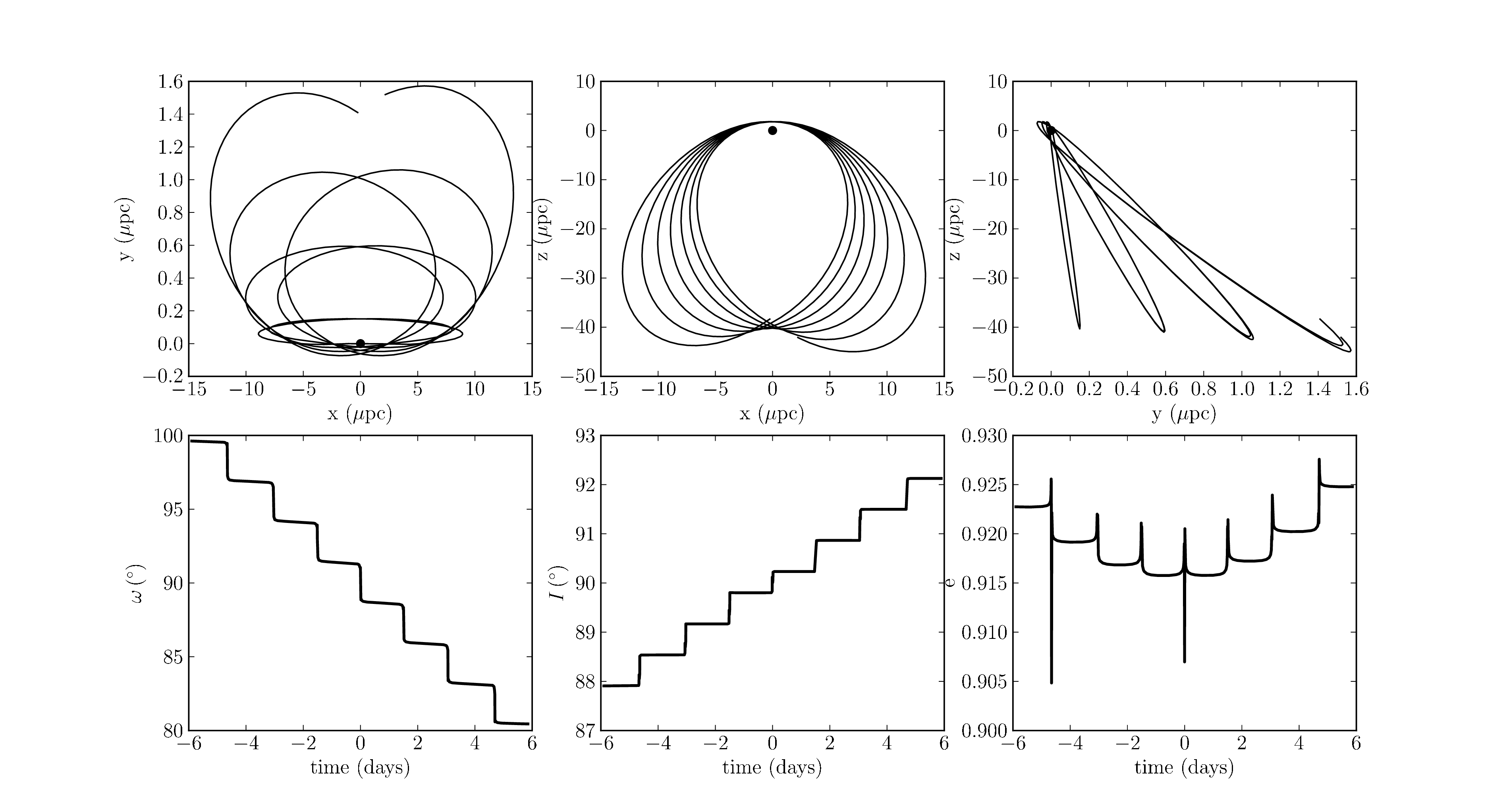}
\caption{Orbital effects of $H_{s^2}$ (even-spin terms).  The orbit is
like a 1/300-size version of S2 ($a=0.02\,$mpc and $e=0.88$), initially in the $x$-$z$ plane, while
the spin is along $x$. Schwarzschild and frame-dragging terms have been omitted,
so this is a completely artificial example.  It is nonetheless
interesting, as it illustrates the complexity of the spin-derived
effects, from which none of the Keplerian orbital elements are exempt
from change.} \label{fig:orbit_H6}
\end{figure*}

Figures~\ref{fig:orbit_H4} and \ref{fig:orbit_H5}, mentioned in the previous section, show the orbit effects $H_{\rm sch}$ and $H_{s}$ as a perturbation to pure Newtonian motion. Cumulative precession is a discrete phenomenon which occurs at pericentre. However, the effect is not completely step-like, with transient behaviour before and after the pericentre kicks.
The complicated $H_{s^2}$ orbit evolution effects are shown in Figure~\ref{fig:orbit_H6}.
The evolution depends on the relative orientation of the orbital angular momentum with the black hole spin. 
We do not have any
interpretation that helps understand the dynamics generated by these
higher-order terms, and merely show this orbit as an example.  

\begin{figure*}
\includegraphics[scale = 0.6]{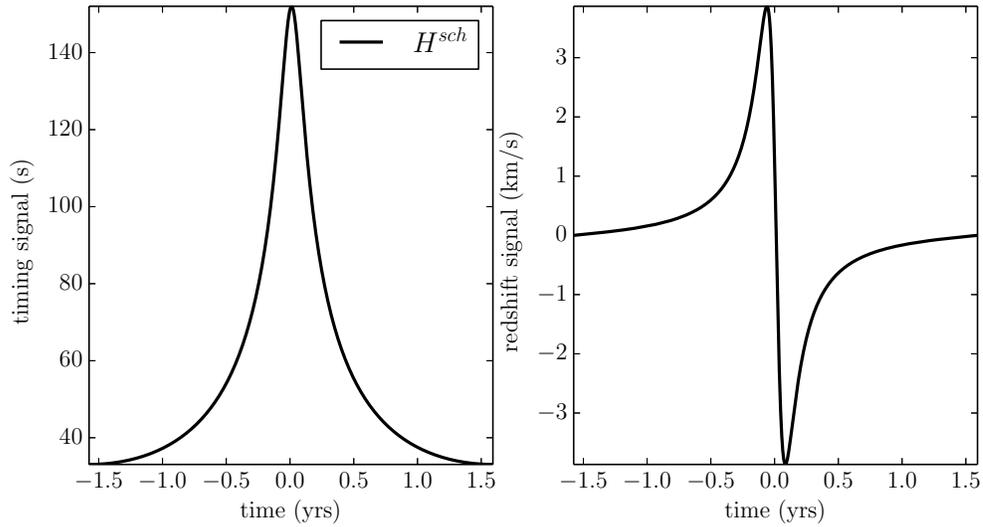}
\caption{Light-path contributions to time delays and redshifts of
Schwarzschild terms. The left panel is simply the well-known Shapiro
delay. The redshift is the derivative of the time-delay, stated in eq. (\ref{crux}). The orbital period is $3\,$yrs - a scaled version of the orbits used for Fig. ~\ref{fig:funclox}: $e=0.6$, $a=0.041''$, $I=45^\circ$, and $R_0=8.31\,$kpc.  The time delay depends only logarithmically
on the period $P$, but the redshift signal scales as $1/P$. Because this is a light-propagation effect, there is no cumulative component to this effect. }                 
\label{fig:signalsSCH}
\end{figure*}

\begin{figure*}
\includegraphics[scale = 0.5]{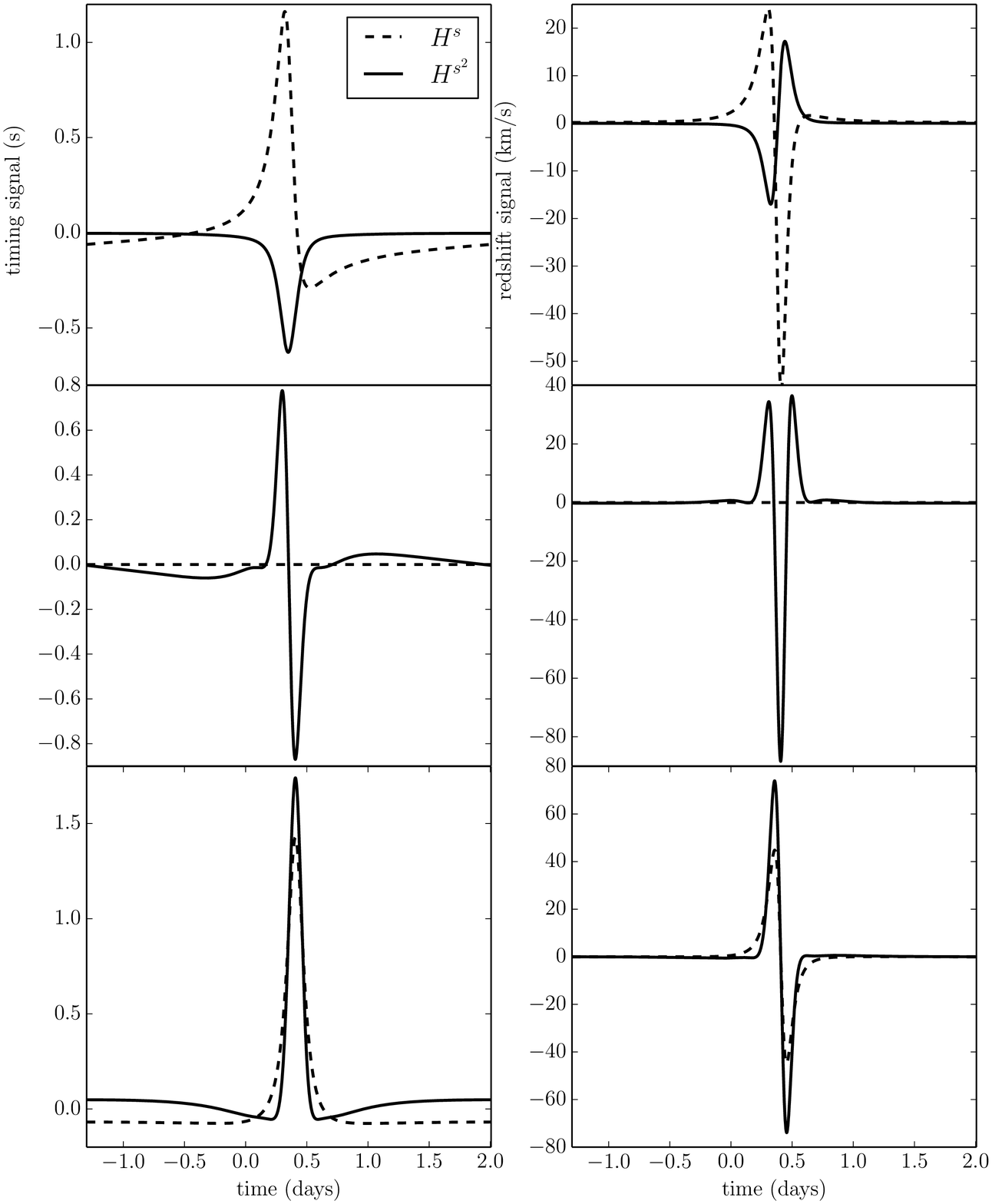}
\caption{Light-path contributions of spin terms to time delays and
redshifts.  The orbit in each case is 1/50-sized version of the one in
Figure~\ref{fig:signalsSCH} ($a = 0.0008''$, $e = 0.6 $), while the spin is maximal.  Only the spin
direction changes: in the top row, the spin is perpendicular to the
orbit; in the middle row, the spin is along the line of sight, and
hence the frame-dragging signal is null, leaving only even-spin
contributions; in the bottom row, the spin is perpendicular to the
line of sight.} \label{fig:signalsSpin}
\end{figure*}

Moving now to light-path effects, Figure~\ref{fig:signalsSCH} shows
the contribution of $H^{\rm sch}$, and Figure~\ref{fig:signalsSpin}
shows the contributions of $H^s$ and $H^{s^2}$.  The even-spin signals on timing and
redshift are capable of a wide variety of signal shapes, which depend on
the orbit geometry relative to the observer and the spin-direction. Timing delays due to spin effects influencing photon paths have been calculated for binary pulsars \citep[see for example Fig. 5 in][]{0004-637X-514-1-388}. 

\subsection{Scaling}
Table~\ref{tab:Terms} also gives the scaling of the time delay, which
depends on some power of the orbital period $P$.  For the classical
Kepler or R\o mer effect $\Delta t \sim P^{2/3}$.  With respect to the
relevant terms in the Hamiltonian, we thus have
\begin{equation}
   \mathbf{p}^2
   \; \Rightarrow \; \Delta t \sim P^{2/3} \,.
\end{equation}
The $p_t^2/r$ behaves differently for orbits and light paths. For
orbits, it is of course part of Keplerian dynamics.  For light paths
it is part of the Shapiro delay, which depends only logarithmically on
$r$.  Accordingly, we write
\begin{equation}
   \frac{p_t^2}r \; \Rightarrow \; \Delta t \sim
   \begin{cases}
      P^{2/3}  & \mbox{orbits}      \\
      P^0      & \mbox{light paths}.
   \end{cases}
\end{equation}
Table~\ref{tab:Terms} also has many terms which look like
increasingly elaborate versions of the classical ones.  The scaling of
$\Delta t$ for such terms is simple: provided we are not close to the
horizon, a factor of $1/r$ in a Hamiltonian term introduces a factor
$P^{-2/3}$ in the time delay.  That leaves only the 
$p_t\,\bf{x} \times \bf{p}$ term to deal with.  To do that, we consider the
geometric mean of $\mathbf{p}^2$ and $p_t^2/r^4$ to get
\begin{equation}
   \frac{p_t \bf{x} \times \bf{p}}{r^3} \; \Rightarrow \; \Delta t \sim
   \begin{cases}
      P^{-1/3}  & \mbox{orbits}             \\
      P^{-2/3}  & \mbox{light paths}.
   \end{cases}
\end{equation}
Table~\ref{tab:Terms} includes all terms with $\Delta t$
contributions up to $P^{-2/3}$.

Redshifts scale like
\begin{equation} \label{division}
   \frac{\Delta t}P
\end{equation}
as will be evident from Equations~(\ref{crux}) and (\ref{crux2}).
This assumes, as before, that orbits and light paths are not too close
to the horizon.  This suggests that prospects for testing relativity
as period sizes decrease improve quicker for stellar orbits than
pulsar orbits. 

\begin{figure}
\includegraphics[scale = 0.5]{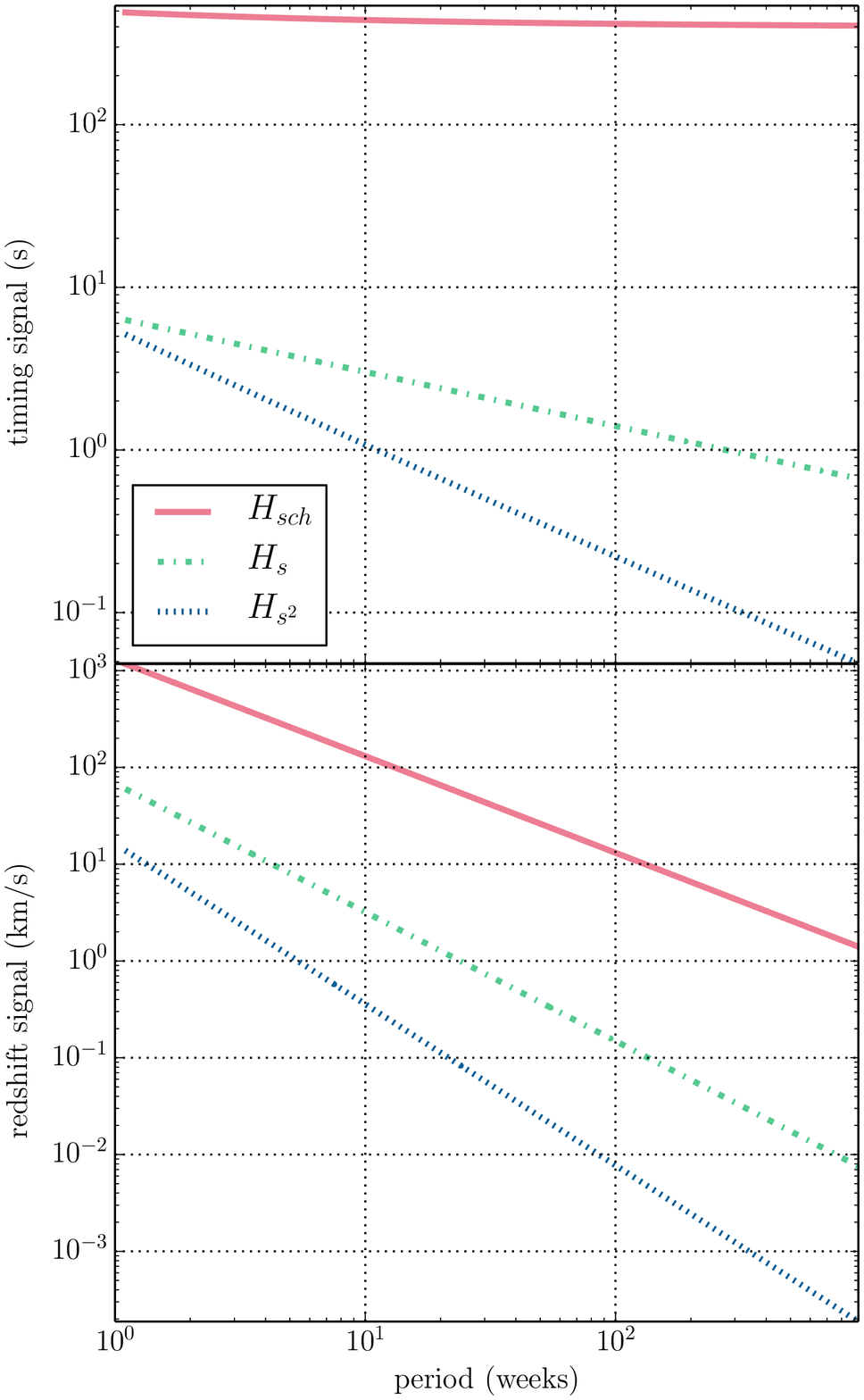}
\caption{Transient orbital contributions to time delays and redshifts of
different relativistic terms (Schwarzschild, frame-dragging and spin-squared),
as a function of orbital period. The cumulative components of these relativistic effects have significantly smaller amplitude. The orbit size used in
Figure~\ref{fig:funclox} corresponds to the short-period end of these
panels. The orbital geometry we maintain over this calculation has $I=45^\circ$, and $e=0.6$. }
\label{fig:orbit_scaling}
\end{figure}

\begin{figure} \includegraphics[scale = 0.5]{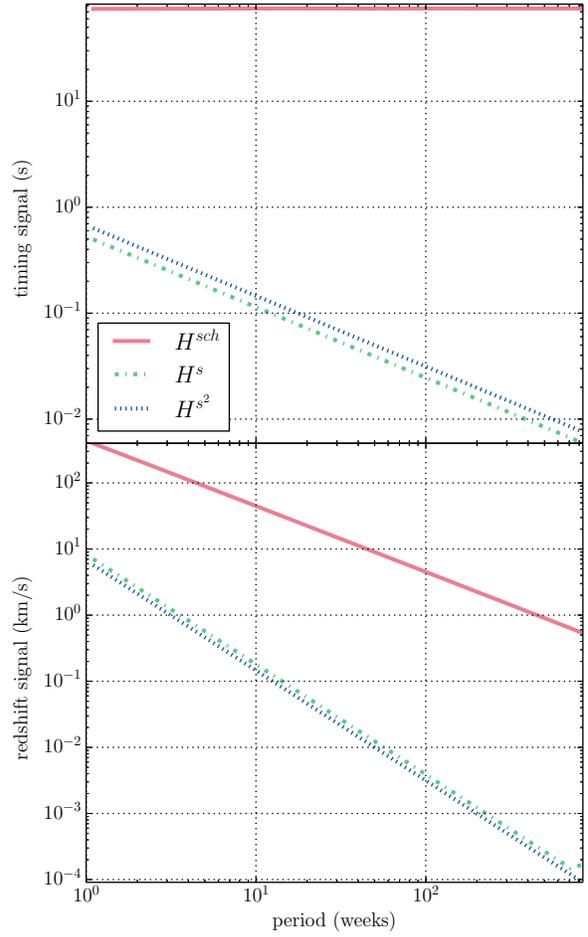}
\caption{Light-path contributions to time delays and redshifts of
different relativistic terms.  This figure complements
Figure~\ref{fig:orbit_scaling}.  The main difference is in how the
odd-spin (or frame-dragging) term scales.}
\label{fig:propagation_scaling} \end{figure}

We can test these scalings with numerical experiments. In Figure~\ref{fig:orbit_scaling}, we show how the transient-relativistic 
contribution of $H_{\rm sch}$, $H_s$ and $H_{s^2}$ scale with the
orbital period $P$. We isolate the transient signal by calculating the most-positive plus most-negative difference in the observables, upon
initialising two orbits at pericentre integrating over one period with
and without the Hamiltonian terms in question. All orbits have
$e=0.6,I=45^\circ$, as in Figure~\ref{fig:funclox}, but the period was
varied.  As we can see in the figure, the predicted scalings from
Table~\ref{tab:Terms} are borne out.  Note, in particular, that the
leading-order Schwarzschild effects on the orbit makes for timing
signals which remain constant as the orbital size deceases.  The
redshift contribution of Schwarzschild however, scales as $1/P$.
Figure~\ref{fig:propagation_scaling} then shows how the light-path
contributions scale with period.  For the latter figure, the spin is
maximal and perpendicular to the orbit, but this detail is unimportant
for the scaling.

For both orbit and light-path effects, our simulations show that
\begin{enumerate}
\item[(i)] the relativistic contributions are concentrated around pericentre, and
\item[(ii)] vary along the orbit in a complicated way, (especially when spin is included), yet
\item[(iii)] nonetheless agree with the orbital period scalings in Table~\ref{tab:Terms}.
\end{enumerate}

As we shall see in the next section, (i) and (iii) will prove useful for extracting relativistic signals from extended mass noise.

\FloatBarrier

\FloatBarrier

\section{Filtering Newtonian perturbations}
\label{sec:wavelets}

Orbit fitting in the pure Kerr case poses no fundamental problems \citep{paper2}, however, critical to being able to resolve relativistic effects on galactic centre stars will be the handling of other perturbations. The most significant are expected to be those from the extended mass distribution, mainly from other stars, but also perhaps from a significant dark matter component. \cite{Merritt}, \cite{Antonini} and \cite{iorio} compare the cumulative effects of extended mass and relativity.

In this section we are interested in transient relativistic signals over a single orbit. A star whose redshift/time-delay is expected to be influenced by relativity is the target star. The redshift/time-delay of this star is also affected by the Newtonian attraction of other black hole-orbiting stars in the neighbourhood, which we call the perturbers.

While the relativistic time dilation signal is likely to be stronger
than extended Newtonian signals, the next-strongest effects
(Schwarzschild and Shapiro) may be partially obscured. In this section
we first discuss how to calculate the Newtonian perturbations on the
target star, before introducing a wavelet decomposition method as a
tool which could be used to help distinguish them from relativistic
perturbations.

\begin{figure}
\includegraphics[scale = 0.6]{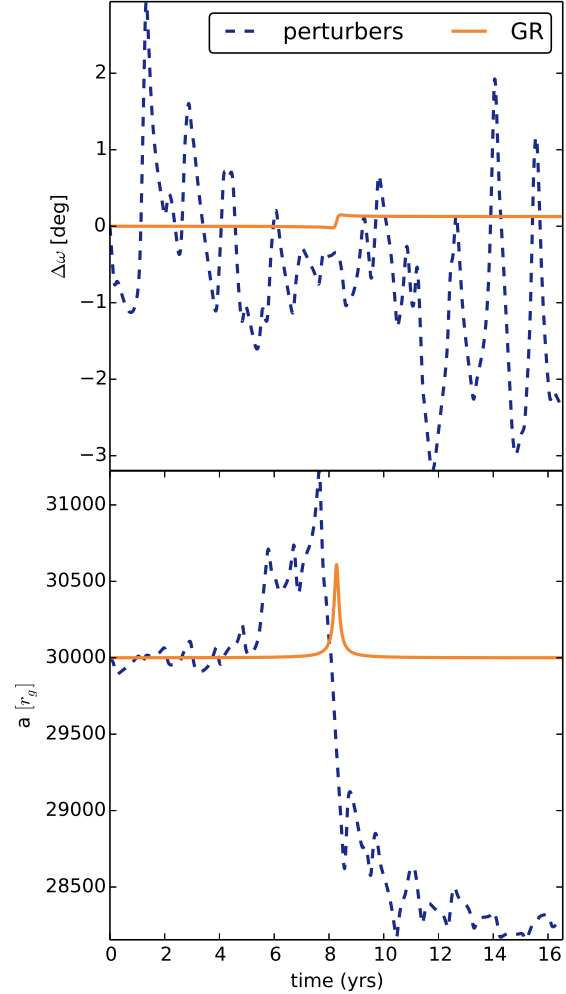}
\caption{The evolution of the instantaneous Keplerian elements $a$ and $\omega$. The Newtonian perturbations from other stars are distributed throughout the orbit, whereas relativistic perturbations are concentrated near pericentre.  (Note that the instantaneous $a$ and $\omega$ are not directly observable in relativity, because they are gauge-dependent and don't take signal propagation into account.  The observable quantities are arrival times and redshifts.)}
\label{fig:element_movement}
\end{figure}

\subsection{Newtonian perturbers}

The classical leading-order perturbation due to other stars orbiting
the black hole is given by a Hamiltonian contribution
\begin{equation}\label{eq:wisdom}
H_\textrm{stellar} = \sum_{j} \frac{m_j}M
\left( \frac{{\vec x} \cdot {\vec x}_j}{|{\vec x}_j|^3}
     - \frac1{|{\vec x}-{\vec x}_j|} \right),
\end{equation}
where ${\vec x}$ is the star being observed and $m_j,{\vec x}_j$ refer
to perturbing stars.  For a derivation, see \cite{wisdom}, especially
their equation (17), and disregard the mutual perturbations of the
${\vec x}_j$ stars.  Note however, that the back-reaction on the
observed star due to the perturbed position of the black hole must be
included.  We model the perturbations by adding the classical
perturbation \eqref{eq:wisdom} to the relativistic Hamiltonian from
Table~\ref{tab:Terms}.  \cite{PhysRevD.89.044043} shows that new
relativistic terms appear in general $N$-body problems, if there is a
tidal force or a quadrupole of the same order as the dominant
monopole.  If the star being observed were in a binary, such terms
would arise, but for the simpler problem we are considering, the
approximation of simply adding the classical perturbers appears to be
valid.

As an example of the effect of Newtonian perturbers, we consider a
target star at $\vec x$ on an S2-like orbit \citep{gillessenMonitors}
with semi-major axis $a = 30000$ in geometric units, and eccentricity $e=0.9$. The perturbers
at ${\vec x}_j$ are 100 stars, all of equal mass, together making up
$1\%$ of the black hole's mass or $\simeq 4\times10^4M_\odot$.  These
are distributed according to a power-law profile $\rho\left( r\right)
\propto \exp\left({-\gamma r}\right)$ with $\gamma = 0.5$. Their
eccentricity distribution is uniform.  Figure
\ref{fig:element_movement} contrasts the Newtonian and relativistic
perturbations on the target star's semi-major axis $a$ and periapsis
argument $\omega$.  As we see, the relativistic perturbations are
completely submerged under the Newtonian stellar perturbations.  We
may recall that for Mercury, Newtonian perturbations from other masses
are an order of magnitude larger than the relativistic effects, yet
the accumulation of $\Delta\omega\simeq0.1''$ per orbit is measurable.
What makes such a measurement possible is that in the solar system,
planetary masses are known accurately and hence the Newtonian
perturbations can be subtracted off.  Near the Galactic Centre, there
is no prospect of measuring all the perturbing masses
accurately. Hence, if the model perturbers in Figure
\ref{fig:element_movement} are at all representative, relativistic
effects would be drowned under Newtonian perturbations.

\subsection{Wavelets}

However, the situation is not hopeless.  Because the transient
relativistic effects have a very specific time dependence that is
known in advance, it may be possible to extract them from under the
Newtonian background.  Matched-filter techniques, well known from
gravitational-wave searches \citep[see for
  example][]{2009LRR....12....2S}, will not work because the
observables are non-linear in the perturbing effects.  But progress may
be possible using wavelets.

\begin{table}
\caption{The coefficient structure of a wavelet-transformation.  The
  first index gives the time scale, the second index give the
  localisation.}\label{tab:wavelettable}
\begin{center}
\begin{tabular}{c| l}
$n$ & $C_{n,m} $ \\ \hline
$0$ & $C_{0,1}$ \\
$1$ & $C_{1,1}$ \\
$2$ & $C_{2,1}, C_{2,2}$ \\
$3$ & $C_{3,1}, C_{3,2}, C_{3,3}, C_{3,4}$\\
$4$ & $C_{4,1}, \hspace{5mm}\ldots  \hspace{5mm},C_{4,8}$ \\
$5$ & $C_{5,1}, \hspace{5mm}\ldots  \hspace{5mm},C_{5,16}$ 
\end{tabular}
\end{center}
\end{table}

\begin{figure}
\includegraphics[scale=0.68]{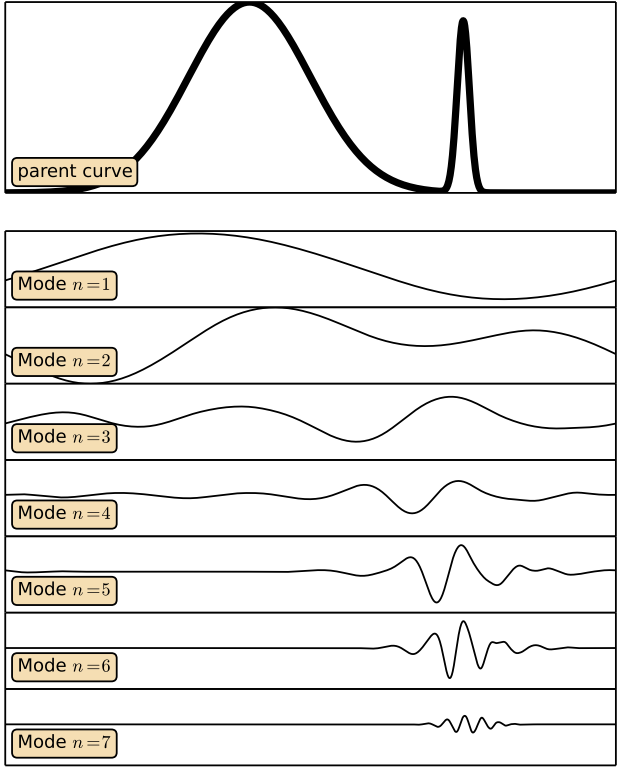}
\caption{In this demonstration, a sum of two Gaussians (top panel) is
  decomposed into the Daubechies~20 wavelet basis (other panels).
  Each of the wavelet panels corresponds to \eqref{eq:Wnoper} for the
  stated value of $n$. (The constant $0$-th mode is not shown.) The
  sum of the lower panel curves yields the parent
  shape.}\label{fig:demo}
\end{figure}

A wavelet decomposition \citep{wavelet1, wavelet2}, allows one to
identify features by breaking down a signal according not just to the
frequency at which they occur, but also according to the time they
occur.  In contrast to a Fourier decomposition, where each basis
function carries frequency information only, a wavelet basis function
includes both frequency and localisation information. Relativistic
perturbations and perturbations due to the extended mass affect the
dynamics in different ways, as Figure~\ref{fig:element_movement}
illustrates, at different frequencies and different localisations.  We
are interested in designing a procedure which helps identify
relativistic signals when shrouded by significant extended-mass noise.
Because redshift curves over a single orbit have no periodicity, and
because relativistic perturbations are most prevalent around
pericentre, wavelets are a natural choice for designing
filters. As a result of relativistic effects being most pronounced
around pericentre --- and non-lingering due to their oft transient
nature --- we can expect high-frequency coefficients, localized around
pericentre passage, to be of greatest value in retaining information
from relativistic effects.  We would expect the extended mass
perturbations to also impart transient, high-frequency effects, such
as close encounters, but those would not be concentrated around
pericentre.

\begin{figure}
\includegraphics[scale = 0.6]{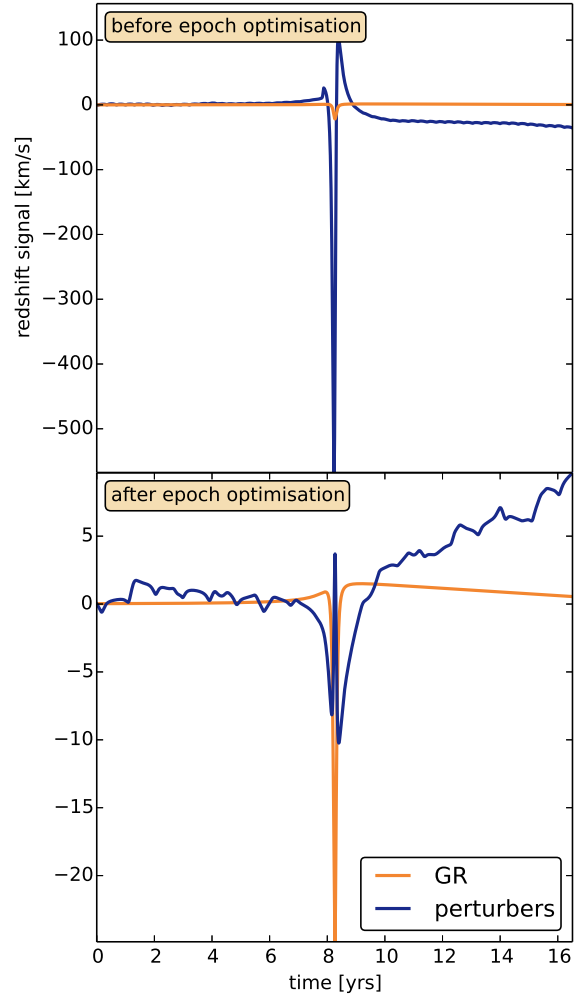}
\caption{The upper panel shows the pure signals $z_{\textrm{GR}} -
  z_{\textrm{Kepl}}$ and $z_{\textrm{Pert}} -z_{\textrm{Kepl}}.$ In
  our chosen example, the perturbation from the extended mass
  distribution is $\sim 30$ times larger than the relativistic
  signal. The lower panel shows the signals after shifting epochs to
  minimise the area under the curves. It is the latter we plug into
  the wavelet procedure detailed in section
  \ref{sec:wavelets}.} \label{fig:epoch_optimisation}
\end{figure}

In a typical wavelet decomposition, such as the Daubechies~4 and
Daubechies~20 wavelet types, a signal is expressed as
\begin{equation}
z(t) = \sum_n \sum_{i=1}^{2^n} C_{ni} \psi_{ni}(t).
\end{equation}
The wavelet basis functions $\psi_{ni}(t)$ are the scaled and
translated versions of a single function, called the mother wavelet,
while the $C_{ni}$ are the expansion coefficients.
Table~\ref{tab:wavelettable} schematically outlines the wavelet
coefficient structure.  Each row of this table corresponds to a
particular time scale, which is twice as fast as in the row above it.
The $n$-th row has $2^n$ coefficients, each of which correspond to
different time windows (or localisations). Let us write
\begin{equation} \label{eq:Wnoper}
\mathcal{W}_n\, z(t) \equiv \sum_{i=1}^{2^n} C_{ni} \psi_{ni}(t).
\end{equation}
The operator $\mathcal{W}_n$ isolates a particular time scale in the
signal.\footnote{We will speak of wavelet frequencies in this section,
  even though we really mean time-scalings of the wavelets.}

Figure \ref{fig:demo} shows the result of applying the $\mathcal{W}_n$
operator to an example curve, consisting of two superposed Gaussians
with different means and widths.  We see that $n=2$ the first (wider)
Gaussian dominates, at $n=3$ the second Gaussian starts to take over,
and from $n=5$ the first Gaussian has been completely filtered out and
only the narrower Gaussian contributes.

\subsection{Filtering relativistic signals with wavelets}

We now consider an S~star (or S~pulsar) whose redshift (or pulse
arrival times) are contaminated by significant noise from an extended
mass system, and investigate how the wavelet coefficients are
influenced by relativistic versus extended-mass perturbations.
Starting with an unperturbed Keplerian orbit, we proceed as follows.

First, we generate three redshift curves for this orbit:
$z_{\textrm{Kepl}}$ has no perturbations, $z_{\textrm{Pert}}$ includes
the Newtonian perturbation by including the effects of
\eqref{eq:wisdom}, and $z_{\textrm{GR}}$ includes only the
relativistic Schwarzschild and Shapiro perturbations. The differences
$z_{\textrm{GR}} - z_{\textrm{Kepl}}$ and $z_{\textrm{Pert}}
-z_{\textrm{Kepl}}$ are plotted in the upper panel of Figure
\ref{fig:epoch_optimisation}. Here we use the same extended Newtonian
mass system example as earlier in this section, corresponding to
Figure~\ref{fig:element_movement}. In this mock data example, the
relativistic redshift signal is $\sim 30$ times weaker than that due
to the extended mass perturbations.

Before taking wavelet transforms, another step is necessary: we need
to choose the reference orbit $z_{\textrm{Kepl}}$ anew, because of
course the ``original'' unperturbed orbit will not be provided by
data.  It would be natural to choose a reference orbit that best fits
the data, but any consistent convention can be used.  For simplicity,
we shift the epoch of $z_{\textrm{Kepl}}$ so as to minimize the
integrated difference from $z_{\textrm{GR}}$ and $z_{\textrm{Pert}}$
respectively.  We denote the shifted Keplerian curves as
$\tilde{z}_{\textrm{Kepl}}$ and $\bar{z}_{\textrm{Kepl}}$. The
differences $z_{\textrm{GR}}-\tilde{z}_{\textrm{Kepl}}$ and
$z_{\textrm{Pert}}-\bar{z}_{\textrm{Kepl}}$ are plotted in Figure
\ref{fig:epoch_optimisation}'s second panel. In this example, they
have approximately the same amplitude.

\begin{figure*}
\includegraphics[scale = 0.55]{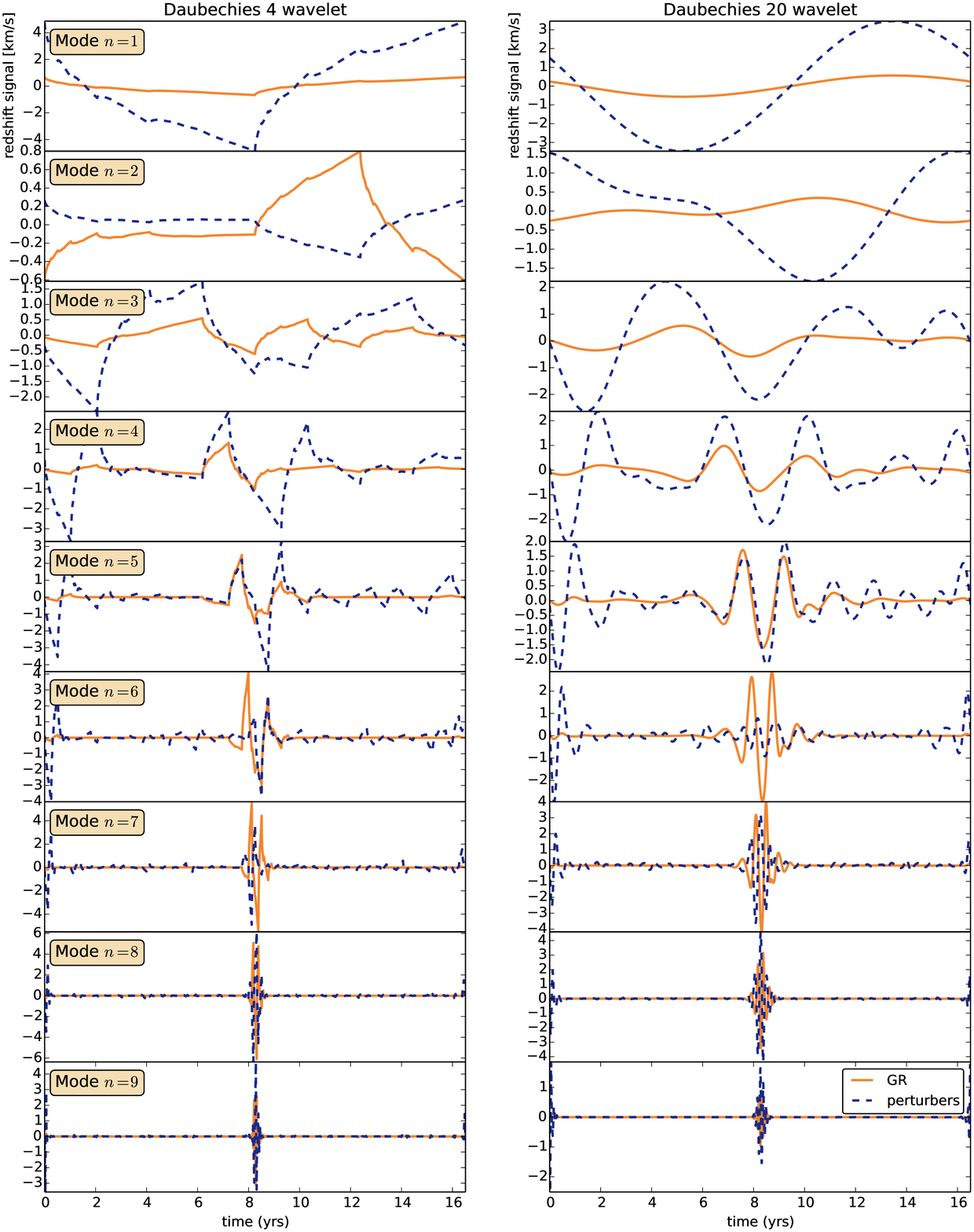}
\caption{Reconstructed signals from wavelet frequency modes. Due to
  the linearity of the wavelet transformation, the sum of the signals
  in each column yield the lower panel curves of figure
  \ref{fig:epoch_optimisation}. The wavelets down the left column are
  the Daubechies 4 variety, and those down the right the Daubechies
  20. Despite the raw relativistic signal being 30 times less than
  that from Newtonian effects due to the perturbing stars, the
  relativistic signal manages to significantly dominate at $n =
  6$. This happens for both wavelet types used here, Daubechies 4 and
  Daubechies 20. This procedure highlights the specific time and
  frequency localisation properties of the two effect types, and tools
  like this could aid future searches in decorrelating
  them.} \label{fig:frequency_slices}
\end{figure*}

We then decompose the signals into different modes according to
frequencies and plot the differences
\begin{equation}
\mathcal{W}_n\, \left( z_{\textrm{GR}} - \tilde{z}_{\textrm{Kepl}} \right)
\end{equation}
and 
\begin{equation}
\mathcal{W}_n\, \left( z_{\textrm{Pert}} - \bar{z}_{\textrm{Kepl}} \right)
\end{equation}
for $n = 1\ldots 9$. These are plotted in Figure (\ref{fig:frequency_slices}), which shows the results using two different wavelet basis functions. 

In our example, the wavelet-reconstructed perturber signal is stronger
than the relativistic ones over all wavelet scales, except at $n=5$
(with twice the amplitude) and $n=6$ (with almost the same amplitude).
This decomposition procedure indicates that given the geometry of our
chosen orbit, Schwarzschild effects, although obscured by extended
mass perturbations with a signal-to-noise $S/N \sim 1/30$, will impart
a significant contribution on the $n=6$ frequency modes.
Alternatively, one can say that Schwarzschild effects, though about
30-fold weaker overall than extended-mass perturbations (in this
model), nonetheless stand out over Newtonian perturbations over a
two-year interval around pericentre, in wavelet modes of timescale
$\approx6$~months.

The above suggests that subtracting off a Keplerian orbit, applying a
wavelet transform to the residual, and then considering a specific
subset of the wavelet coefficients may succeed in filtering out
Newtonian perturbations.

\section{Conclusions and Outlook}

Galactic centre stars travel upon the most relativistic orbits
known. However, the accessible relativistic effects are not simply
extensions of similar experiments in the solar system and in binary
pulsars.  S~stars, and S~pulsars if they exist, live in stronger
fields than binary pulsars, but their orbital periods are much longer.
The combination of the strong fields, long orbital time-scales, and
the typically high eccentricities push otherwise negligible aspects of
dynamics near a black hole into the observables.  For example,
precession is not a steady process, as the well-known orbit-averaged
formulas (\ref{mercury}) and (\ref{fd_precession}) may suggest, but
nearly a shock that happens at pericentre.  The concentration of
dynamical effects around pericentre passage applies even more to
effects which depend on the spin of the black hole. These pericentre
shocks will be important when separating relativistic signals from
noise sources.

If the observed dynamics is found to be in agreement with a Kerr
space-time plus perturbations from the surrounding astrophysical
environment, Einstein gravity will be tested to a new level. A further
benefit is that because we test gravity by tracking freely-falling
bodies, as well as photon paths, by inferring the components of the
metric by looking at their effects on the behaviour of geodesics, we
probe not just the field equations, but implicitly test the notion requisite to describing gravity with
geometry - the principle of equivalence.

The spectrometers of the Keck and VLT telescopes have independently
observed the spectra of the S~stars, managing to achieve spectral
resolution to $\sim 10\rm\,km\,s^{-1}$ in the best cases.  The next
generation of instruments, such as the High Resolution Near-infrared 
Spectrograph (SIMPLE) on the E-ELT is expected to
achieve $\sim 2\rm\,km\,s^{-1}$ \citep{simple}. If an S~star with period $\sim 1$ year is discovered, observations clustered around pericentre passage at this level of accuracy could provide a measurement for frame-dragging. Were S~pulsars with stable periods to be detected with orbits similar
to the already-known S~stars, pulsar timing even at the msec level
would be, in principle, enough for all the effects summarised in
Table~\ref{tab:Terms}.  The challenge would be removing the Newtonian
``foreground'' due to the extended mass distribution around Sgr~A*.
Separating cumulative effects into Newtonian versus relativistic is
a challenging task, yet with transient effects that vary along an orbit
in different ways, one can be more optimistic.

\FloatBarrier

\section*{Acknowledgements}
We thank S. Gillessen and S. Tremaine for many helpful insights and B.P. Schmidt for some suggestions. We thank the referees, whose comments have helped to improve the paper. R.A. acknowledges support from the Swiss National Science Foundation. 
\bibliographystyle{mn2e}
\bibliography{clocks_around_SgrA}

\end{document}